\documentclass[twocolumn,prl,floatfix,superscriptaddress,nobibnotes,nolongbibliography]{revtex4-2}

\usepackage{amsmath}
\usepackage{amssymb}
\usepackage{graphicx}
\usepackage{hyperref}
\usepackage[normalem]{ulem}

\usepackage{xcolor}

\hypersetup{
    colorlinks=true,
    linkcolor={blue!80!black},
    citecolor={blue!80!black},
    urlcolor={blue!80!black}
}

\usepackage{color}
 
\begin{document}

\title{Caroli--de Gennes--Matricon Analogs in Full-Shell Hybrid Nanowires}

\author{M.~T. Deng}
\affiliation{Center for Quantum Devices, Niels Bohr Institute, University of Copenhagen, 2100 Copenhagen, Denmark}
\affiliation{Institute for Quantum Information \& State Key Laboratory of High Performance Computing, College of Computer Science and Technology, NUDT, Changsha 410073, China}
\author{Carlos Pay\'a}
\affiliation{Instituto de Ciencia de Materiales de Madrid (ICMM), CSIC, 28049 Madrid, Spain}
\author{Pablo San-Jose}
\affiliation{Instituto de Ciencia de Materiales de Madrid (ICMM), CSIC, 28049 Madrid, Spain}
\author{Elsa Prada}
\affiliation{Instituto de Ciencia de Materiales de Madrid (ICMM), CSIC, 28049 Madrid, Spain}
\author{C. M. Marcus}
\affiliation{Center for Quantum Devices, Niels Bohr Institute, University of Copenhagen, 2100 Copenhagen, Denmark}
\affiliation{Materials Science and Engineering, and Department of Physics, University of Washington, Seattle WA 98195}
\author{S. Vaitiek\.{e}nas}
\affiliation{Center for Quantum Devices, Niels Bohr Institute, University of Copenhagen, 2100 Copenhagen, Denmark}

\date{\today}

\begin{abstract}
We report tunneling spectroscopy of Andreev subgap states in hybrid nanowires with a thin superconducting full-shell surrounding a semiconducting core.
The combination of the quantized fluxoid of the shell and the Andreev reflection at the superconductor-semiconductor interface gives rise to analogs of Caroli--de Gennes--Matricon (CdGM) states found in Abrikosov vortices in type-II superconductors.
Unlike in metallic superconductors, CdGM analogs in full-shell hybrid nanowires manifest as one-dimensional van Hove singularities with energy spacings comparable to the superconducting gap and independent of the Fermi energy, making them readily observable.
Evolution of these analogs with axial magnetic field, skewed within the Little-Parks lobe structure, is consistent with theory and yields information about the radial distribution and angular momenta of the corresponding subbands.
\end{abstract}

\maketitle
In moderate magnetic fields, type-II superconductors are threaded by Abrikosov vortices, which confine and quantize the applied flux in units of $\Phi_0 = h/2e$, corresponding to twists of the superconducting phase around the vortex core \cite{Abrikosov1957, Tinkham1996}. Spatial confinement of quasiparticles in the vortex core results in another type of quantization, that of non-dispersive bands of Andreev bound states, referred to as Caroli--de Gennes--Matricon~(CdGM) states~\cite{Caroli:PL64}. 
Energy level spacings of CdGM states are smaller than the superconducting gap, $\Delta$, by a factor of order $\Delta/E_{\rm F}$, where $E_{\rm F}$ is the Fermi energy \cite{Caroli:PL64}. In most conventional superconductors, this factor is typically very small, $\Delta/E_{\rm F} \sim 10^{-4} $, making it difficult to resolve CdGM states experimentally. 
Some signatures of CdGM states have been observed in compound superconductors with lower $E_{\rm F}$ using scanning-tunneling microscopy~\cite{Berthod2017, Chen2018, Chen2020}.
Vortex-core states have been investigated in the context of topological superconductivity, where the spectrum of CdGM states includes a zero-energy mode~\cite{Xu2015, Liu2018a, Wang2018, Kong2019, Ge2023}.

The development of transparent interfaces between semiconductors and superconductors in hybrid nanowires~\cite{Krogstrup2015, Carrad2020, Heedt2021} has provided a new route to tunable confined superconducting modes.
Recent studies have explored bound states in InAs and InSb nanowires with thin superconductors on their surface, mainly in the context of topological superconductivity~\cite{Lutchyn_review2018, Prada2020, Marra_review2022}.
This includes studies of nanowires with a fully surrounding superconducting shell~\cite{Vaitiekenas2020, Vaitiekenas2020_2, Penaranda2020, Kopasov:PRB20, Razmadze2020, Kopasov:PSS20, Sabonis2020, Kringhoej2021, Valentini_Science2021, Vekris:SR21, Valentini:N22, Escribano2022_2, Ibabe:NC23, Razmadze2024, Giavaras:PRB24, Paya2024, Paya2024_2}.
The doubly-connected geometry enables the realization of analogs of CdGM states in the form of van Hove singularities in dispersive one-dimensional (1D) bands~\cite{SanJose2023}.
In the presence of an axial magnetic field, the full shell acquires a phase winding of the superconducting order parameter, resembling a synthetic vortex. 
Due to the fluxoid quantization, the change in vorticity, that is, the number of $2 \pi$ twists of the superconducting phase, is discrete and increases stepwise with flux.
This results in flux-induced oscillations in the superconducting transition temperature, known as the Little-Parks effect~\cite{Little1962}.
The degree to which the superconducting properties get modulated depends on the nanowire diameter, $d$, compared to the superconducting coherence length, $\xi$.
For larger diameters, $d\gg\xi$, scalloped lobes emerge separated by discontinuous jumps in spectral features, marking where one fluxoid enters the core~\cite{Little1962}.
In contrast, for $d \lesssim \xi$ superconductivity is destroyed between lobes~\cite{Liu2001, Schwiete2009, Sternfeld2011}.
We have previously investigated both larger and smaller diameter wires with the shells showing, respectively, nondestructive and fully destructive Little-Parks effects, in both cases agreeing well with theory \cite{Vaitiekenas2020_2}. 

%%%%%%%%%%%%%%%%%%%%%% FIG. 1 %%%%%%%%%%%%%%%%%%%%%%%%
\begin{figure}[t]
\includegraphics[width=\linewidth]{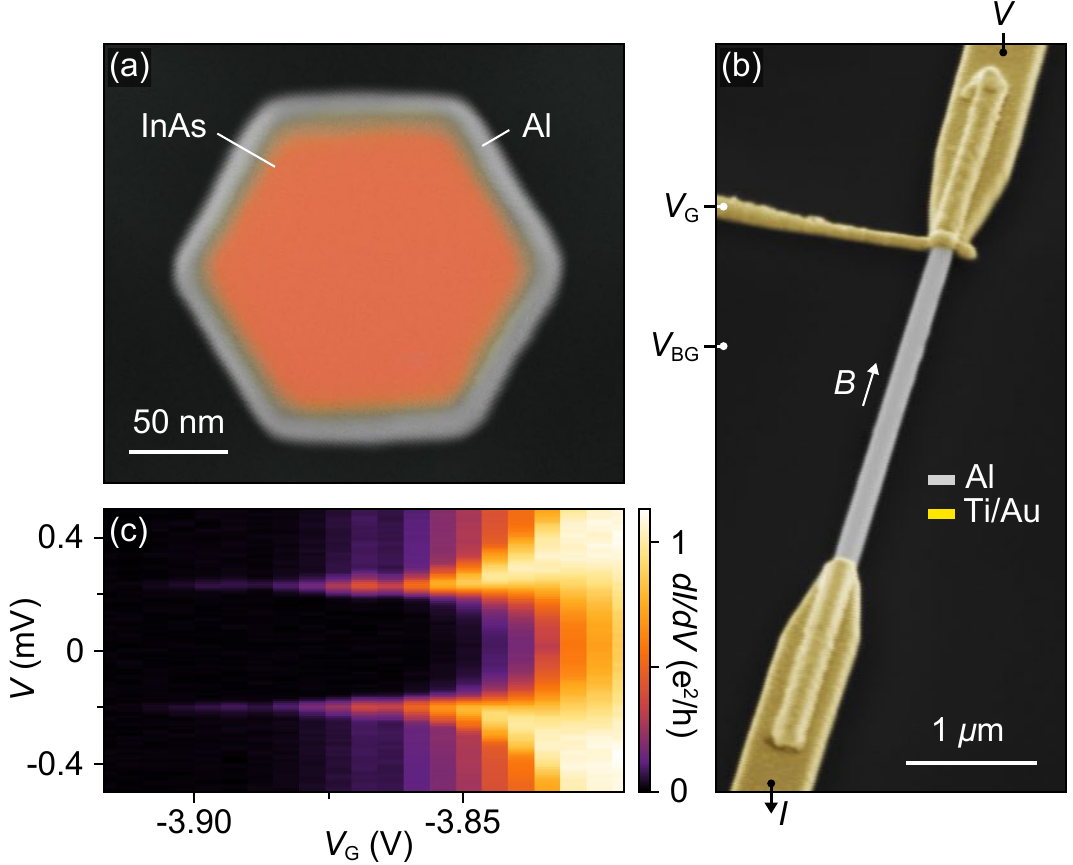}
\caption{
(a) Electron micrograph of a full-shell nanowire cross section from the growth batch of device~1, showing the rounded-hexagon InAs core (orange) and 7~nm Al shell (gray). The image was generated by overlaying separate material-sensitive micrographs.
(b) False-color micrograph of device~1, showing the Al shell (gray) along with Ti/Au contacts (yellow) and wrapping top gate (yellow). Bare InAs under the top gate allows depletion by top gate or back gate.
(c) Differential conductance, $dI/dV$, as a function of voltage bias, $V$, and top-gate voltage, $V_{\rm G}$, at zero magnetic field.
A hard superconducting gap is measured at weak tunneling, $V_{\rm G} < -3.85$~V, with back-gate voltage $V_{\rm BG} = -1$~V.
}
\label{fig:1}
\end{figure}
%%%%%%%%%%%%%%%%%%%%%%%%%%%%%%%%%%%%%%%%%%%%%%%%%%%

In this Letter, we investigate tunneling spectra of CdGM analog subgap states in InAs nanowires with a thin epitaxial Al full shell.
We find that the energies of the CdGM analogs relative to midgap are modulated roughly following the lobed gap of the shell,  $\Delta (B)$. Importantly, however, the CdGM analogs are asymmetric within each non-zero lobe, skewed toward high magnetic field compared to the symmetric lobe edges.
The skewing was explained theoretically in Ref.~\onlinecite{SanJose2023}, where CdGM analogs were modeled using several forms of radial confinement. Here, we show that experimental tunneling spectra are well described by the model, including the skewing, providing detailed understanding of CdGM analogs. 

The devices were fabricated using hexagonal InAs nanowires [Fig.~\ref{fig:1}(a)] from two growth batches with thinner (7~nm) and thicker (24~nm) Al full shells.
Results from two devices, denoted 1 and 2, are presented and show comparable features.
Each device has a normal Ti/Au lead and a $\sim 100$~nm region of bare InAs where the Al has been removed using MF-321 developer as an etchant.
Both devices have a back gate that can be used to deplete the bare InAs region without affecting the shielded wire or contact.
Device 1 has an additional local top gate, patterned over the junction after applying a global atomic layer deposited HfO$_2$ insulator; see Fig.~\ref{fig:1}(b).
The tunneling barrier is formed using the gate electrodes, allowing for spectroscopy from the normal lead into the proximitized InAs core. 
Differential conductance measurements were performed using standard ac lock-in techniques in a dilution refrigerator with a base temperature of 20 mK. 
An axial magnetic field was applied along the wire, as shown in Fig.~\ref{fig:1}(b), using a vector magnet.
For more details about the nanowire growth, the device fabrication, and the transport measurements, see Appendix. 

Gate-voltage dependence of tunneling transport from the wire into the normal lead for device~1 at zero field is shown in Fig.~\ref{fig:1}(c). For weak tunneling, the measured differential conductance, $dI/dV$, is proportional to the local density of states (LDOS) at the wire end~\cite{Tinkham1996}.
The observed hard gap indicates a transparent superconductor-semiconductor interface~\cite{Chang2015}, and the smooth decrease in tunneling conductance, without resonances, indicates a relatively low disorder barrier in device 1.
The absence of subgap features at zero field indicates an absence of unintentional quantum dot states~\cite{Ahn2021} that develop in some devices~\cite{Valentini_Science2021}.

\begin{figure}
\includegraphics[width=\linewidth]{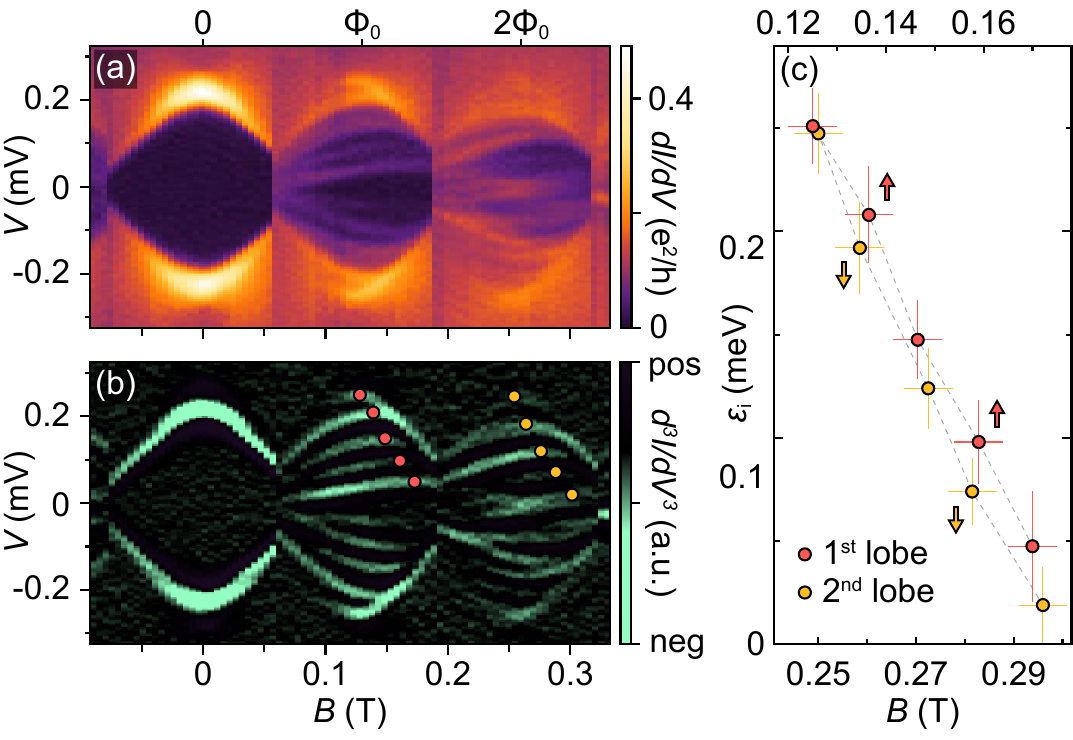}
\caption{\label{fig:2} (a) Differential conductance, $dI/dV$, measured for device 1 in the tunneling regime, as a function of voltage bias, $V$, and axial magnetic field, $B$.
The zeroth lobe shows a superconducting gap without visible subgap states; the first and second lobes show subgap states dispersing with $B$. Data were taken with decreasing $B$.
(b)~Numerical second derivative of the differential conductance, $d^3I/dV^3$, for the data in~(a). Negative (positive) curvature corresponds to a peak (dip) in conductance. Skewness (shift toward larger $B$) is indicated by the shifting positions of subgap-state maxima in the first (red) and second (yellow) lobes.
(c) Subgap-state maxima, $\varepsilon_{\rm i}$, as a function of $B$, measured in the first (top-axis) and the second (bottom-axis) lobes. Skewness is larger for the lower-energy subgap states.
Vertical (horizontal) error bars indicate uncertainties estimated from peak widths ($B$ resolution).
 }
\end{figure}

Tunneling spectrum as a function of the flux-threading axial magnetic field, $B$, shows the Little-Parks modulations with characteristic lobe structure; see Fig.~\ref{fig:2}(a).
In the zeroth lobe, around $B=0$, the gap initially decreases with increasing flux until a discontinuous jump occurs into the first lobe, around the $B$ value where flux reaches a half flux quantum, $\Phi_{0}/ 2$. 
With further increase in flux, similar discontinuous jumps to subsequent lobes are observed. 
The transition points between lobes exhibit a slight hysteresis, likely due to metastable flux states in the superconducting shell.
In contrast to the zeroth lobe, where a hard gap without obvious subgap features was observed, the first and subsequent lobes are filled with discrete subgap states. 
Within each lobe, the subgap states are skewed toward the high-field side of the lobe center.
This effect is highlighted more clearly by plotting the second derivative of conductance with respect to bias voltage, $d^3I/dV^3$, which sharpens the features; see Fig.~\ref{fig:2}(b).
The observed skewness increases systematically with decreasing energy.
We quantify the dependence for the first and second lobes by plotting the energy maxima of each subgap state, $\varepsilon_i$, as a function of $B$, as shown in Fig.~\ref{fig:2}(c), revealing a clear trend also found in the theoretical model~\cite{SanJose2023}, as described below.

We interpret the subgap states as CdGM analogs, which can be ascribed to shell-induced van Hove singularities in 1D core subbands confined radially along the InAs core by the superconducting gap.
The number of twists of the superconducting boundary corresponds to the lobe index. 
This simple picture of 1D CdGM analogs does not take into account the effects of disorder along the wire length, which would lead to full confinement of discrete zero-dimensional Andreev states rather than van Hove singularities in 1D core subbands.
We note that while some disorder along the wire is inevitable, the semiconducting core, fully covered by Al, has not been exposed to air outside of the epitaxy growth chamber.
Therefore, we expect minimal disorder along the wire compared to bare or partially covered nanowires. 

%%%%%%%%%%%%%%%%%%%%%% FIG. 2 %%%%%%%%%%%%%%%%%%%%%%%%
%%%%%%%%%%%%%%%%%%%%%%%%%%%%%%%%%%%%%%%%%%%%%%%%%%%

\begin{figure}
\includegraphics[width=\linewidth]{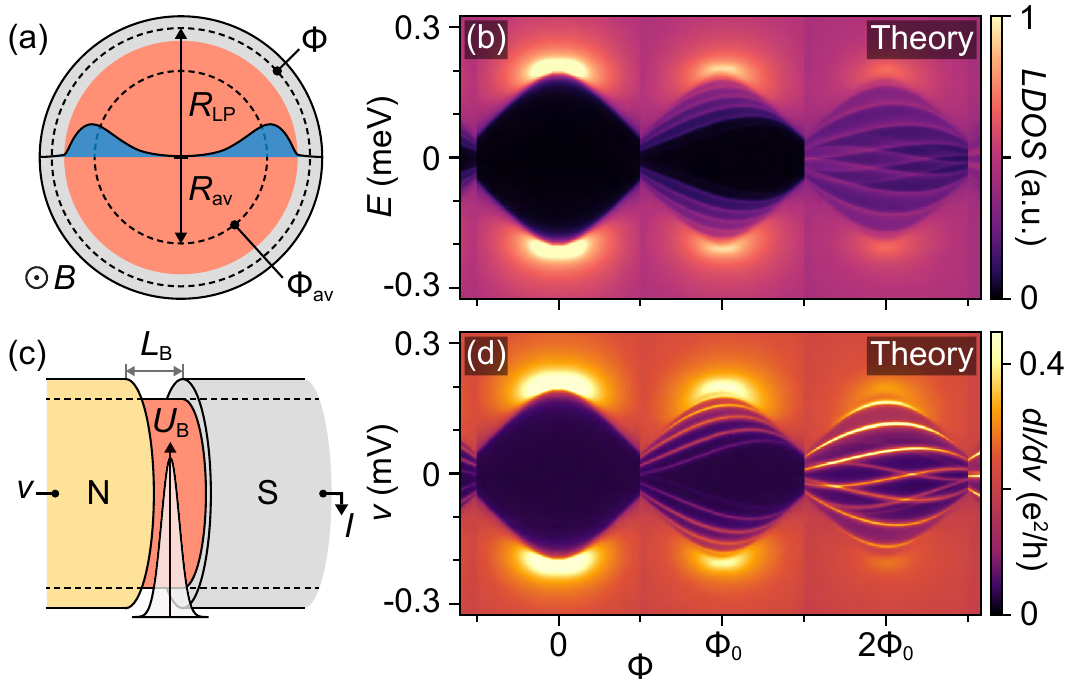}
\caption{\label{fig:3}
(a) Schematic of the full-shell hybrid nanowire cross section in a cylindrical approximation, showing a semiconductor core (orange) fully covered by a thin superconducting shell (grey) with an average radius $R_{\textrm{LP}}$. An applied axial magnetic field, $B$, threads a flux, $\Phi$, through the nanowire. Wave functions of the different CdGM analogs, characterized by generalized angular momentum $m_{\rm L}$, have roughly the same average radius $R_{\textrm{av}}$ enclosing magnetic flux $\Phi_{\textrm{av}}$. 
(b)~Calculated local density of states at the end of a semi-infinite full-shell wire as a function of energy, $E$, and flux, $\Phi$, for the zeroth, first, and second lobes. 
(c)~Schematic of the potential barrier in the uncovered semiconductor region between the normal metal (N) and the full-shell wire (S) with a height $U_{\textrm{B}}$ and length $L_{\textrm{B}}$. 
(d)~Modeled differential conductance, $dI/dv$, through the junction as a function of voltage bias, $v$ (lower case for the model), and flux, $\Phi$. 
Model input parameters (see Ref.~\cite{SanJose2023} for details): $R_{\rm LP} =72$~nm, shell thickness $t=7$~nm, effective mass $m^* = 0.023\,m_e$, zero-field shell gap $\Delta = 0.2$~meV, coherence length  $\xi = 90$~nm, tight-binding parameter $a_0 = 1$~nm, and barrier length $L_{\textrm{B}} = 60$~nm;
fit parameters: $R_{\textrm{av}} = 46$~nm,  Fermi energy $E_{\rm F} = 35$~meV, superconductor-semiconductor coupling $\Gamma = 22\,\Delta$, and barrier height $U_{\textrm{B}} = 70$~meV.}
\end{figure}

To understand the general behavior of the CdGM analogs, including the energy-dependent skewness, we use the modified hollow-core model introduced in Ref.~\onlinecite{SanJose2023}, which is an extension of the Hamiltonian used in Ref.~\onlinecite{Vaitiekenas2020}.
Our simplified model considers a clean cylindrical semiconducting core, as a proxy to the rounded hexagon [Fig.~\ref{fig:1}(a)], covered by a thin superconductor shell, and threaded by a magnetic flux $\Phi=\pi R_{\rm{LP}}^2 B$, where $R_{\rm{LP}}$ is the average shell radius; see Fig. \ref{fig:3}(a).
Charge accumulation from the difference in work functions of the Al shell and InAs core results in a dome-shaped electrostatic potential~\cite{Mikkelsen:PRX18, Antipov:PRX18, Schuwalow:AS21}.
The combination of normal and Andreev reflections at the core-shell interface gives rise to subbands in the core characterized by a generalized angular momentum quantum number, $m_{\rm L}$ \cite{Vaitiekenas2020, Kopasov:PRB20, SanJose2023, Paya2024}. 
A typical resulting wave function, with average radius $R_{\rm{av}}$, is shown in blue in Fig.~\ref{fig:3}(a).

%%%%%%%%%%%%%%%%%%%%%% FIG. 3 %%%%%%%%%%%%%%%%%%%%%%%%
%%%%%%%%%%%%%%%%%%%%%%%%%%%%%%%%%%%%%%%%%%%%%%%%%%%

Calculated local density of states (LDOS) at the end of a semi-infinite full-shell nanowire displaying characteristic lobe structure for the zeroth, first, and second lobes is shown in Fig.~\ref{fig:3}(b).
The modulated gap edge in each lobe is maximal at the center of the lobe, corresponding to an integer multiple of flux quanta threading the area $\pi R_{\rm{LP}}^2$. 
The flux dependence is obtained by solving a self-consistent equation for the pairing amplitude, incorporating flux-induced depairing, and using the results to calculate the LDOS~\cite{SanJose2023}.
The zeroth lobe is empty below the shell gap, while CdGM analogs appear in the first and higher lobes.
The numerical simulations with the chosen wire parameters yield five CdGM analogs for $E>0$ in the first lobe.
The subgap states in the second lobe are similar but have roughly twice larger spread in energy, in some cases resulting in CdGM analogs that cross zero energy.
Within each lobe, the subgap states are theoretically expected to rise in energy linearly in flux, with a slope roughly proportional to $m_L$~\cite{SanJose2023}.
Because $R_{\rm av} < R_{\rm LP}$, the CdGM spectrum is shifted toward higher flux, so the states at lower energy correspond to angular momentum modes with higher $m_{\rm L}$.
At the same time, the slopes are modified by level repulsion from the shell gap, which is most pronounced for the higher energy estates and near the lobe edges. 
The combination of the two effects results in the skewness of the CdGM analogs towards high fields.
As discussed in Ref.~\onlinecite{SanJose2023}, the amount of skewness depends on the ratio $R_{\rm{LP}}^2/R_{\rm{av}}^2$, and reduces to zero (symmetric CdGM analogs) when the two radii coincide (hollow-core approximation).

Modeled differential conductance, $dI/dv$, through a barrier [Fig. \ref{fig:3}(c)] is shown in Fig.~\ref{fig:3}(d). Due to the finite size and height of the barrier, calculated conductance does not strictly follow the density of states.
The modeled spectra reproduce the main features observed experimentally in Fig. \ref{fig:2}(a), including the flux-dependent gap modulation and energy-dependent skewness. 
The good overall agreement with the model, which neglects disorder, suggests that localization effects along the wire do not play a dominant role.
We note that the same model parameters were used in all lobes.

%%%%%%%%%%%%%%%%%%%%%% FIG. 4 %%%%%%%%%%%%%%%%%%%%%%%%
\begin{figure}
\includegraphics[width=\linewidth]{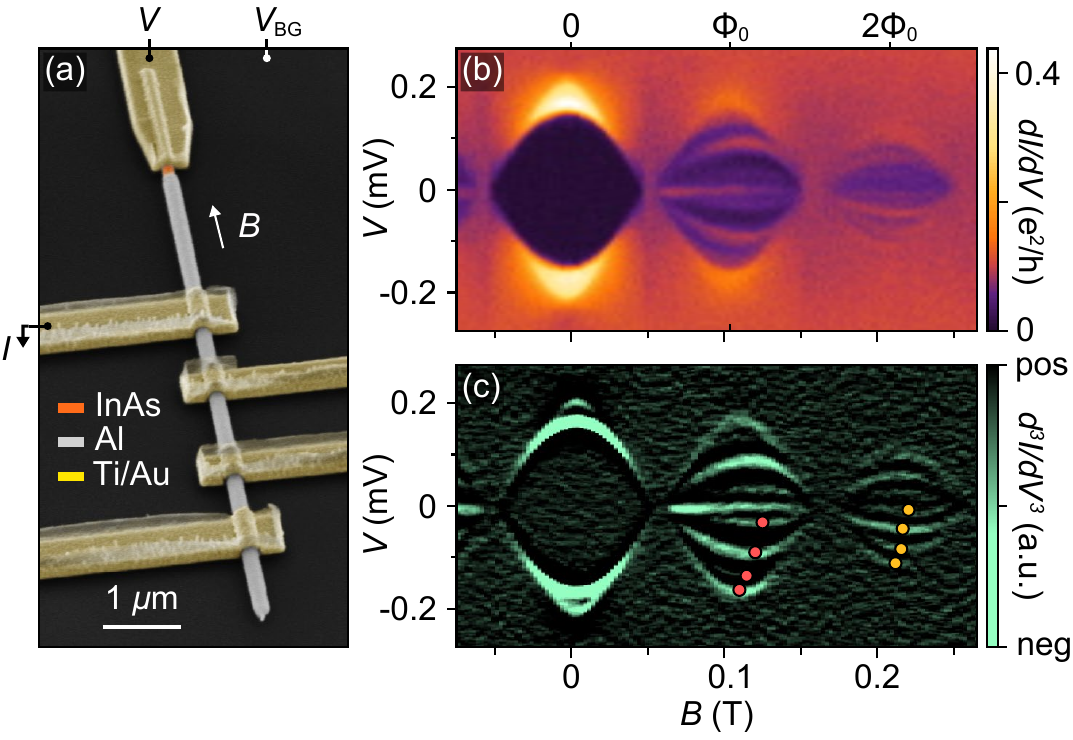}
\caption{\label{fig:4}
(a) Representative color-enhanced micrograph of a thick-shell device with 24~nm Al full shell. The tunneling barrier is tuned with the back gate only.
(b) Tunneling conductance, $dI/dV$, measured for device 2 as a function of source-drain voltage, $V$, and axial magnetic field, $B$.
Similar to the thin-shell device, the zeroth lobe displays a hard superconducting gap, whereas the first and second lobes display several dispersive subgap states.
The first lobe displays a discrete zero-energy state. 
The lobes are separated by featureless normal-state spectra.
(c) The curvature (numerical second derivative) of the conductance, $d^3I/dV^3$, for the data shown in (a). The color points indicate the maxima of each subgap state in the first (red) and second (yellow) lobes.
}
\end{figure}
%%%%%%%%%%%%%%%%%%%%%%%%%%%%%%%%%%%%%%%%%%%%%%%%%%%%%%

Finally, we demonstrate that CdGM analogs can arise in different full-shell nanowires.
The results for device 2, which has a wire from a different growth batch with a thicker Al shell, are summarized in Fig.~\ref{fig:4}. The thicker shell slightly decreases $\Delta$, which in turn increases $\xi \propto \Delta^{-1/2}$~\cite{Vaitiekenas2020_2}.
Furthermore, an enhanced orbital depairing in a thicker shell leads to a steeper overall envelope in $B$~\cite{SanJose2023}.
The combination of these effects results in the destructive regime with a vanishing gap between lobes.
Note that device 2 also shows a zero-energy subgap state along with a typical  spectrum of CdGM analogs. The zero-energy state, not present in device 1, is a characteristic associated with topological superconductivity \cite{Vaitiekenas2020, SanJose2023, Paya2024}.
To better understand the nature of the state would require additional measurements probing its nonlocality~\cite{Prada2020}. 
Experimentally, we find that neither the fully destructive lobe structure nor the presence of a zero-energy state significantly affects the CdGM states, consistent with theory of 1D van Hove states extending along the wire~\cite{SanJose2023, Paya2024}. 

In summary, we have experimentally investigated Caroli--de Gennes--Matricon analogs in InAs nanowires with a fully surrounding epitaxial Al shell. 
The comparable magnitudes of the superconducting gap and the energy spacing of CdGM analogs--set by the semiconductor-superconductor coupling rather than the Fermi energy--enable these states to be readily resolved via tunneling spectroscopy at the end of the nanowire. 
The CdGM analogs were observed in two different wire batches—with and without the destructive regime—suggesting that the phenomenology is independent of shell thickness and core diameter, and is primarily determined by the full-shell geometry.
Good agreement between experiment and theory~\cite{SanJose2023, Paya2024} allows the interpretation of characteristic features, such as skewness of the subgap states, in terms of the angular momentum of 1D CdGM analogs and the reduced average radius of their wavefunction along the semiconductor core.\\

\textit{Acknowledgments}---We thank P. Krogstrup, C.~S\o rensen, and S.~Upadhyay for contributions to material growth and device fabrication. We acknowledge support from research grants (Projects No. 43951 and No. 53097) from VILLUM FONDEN, the Danish National Research Foundation, the European Research Council (Grant Agreement No. 856526), and Grants No. PID2021-122769NB-I00, No. PID2021-125343NB-I00, and No. PRE2022-101362 funded by MICIU/AEI/10.13039/501100011033, “ERDF, EU” and “ESF+”.\\

\textit{Data availability}---All the numerical code used in this Letter was based on the Quantica.jl package \cite{San-Jose:24}. 
The specific code to build the nanowire Hamiltonian and perform the calculations is available in Refs. \cite{Paya:24,Paya:25}.
The data used to generate the figures in this work are available in Ref.~\cite{Deng:25}

\bibliography{bibfile.bib}

\vspace{1em}
\onecolumngrid
\begin{center}
    \textbf{\large End Matter}
\end{center}
\vspace{0.5em}
\twocolumngrid

\textit{Appendix}---\textit{Nanowire growth:} InAs nanowires were grown by molecular beam epitaxy using Au-catalyzed vapor–liquid–solid method on InAs(111)B substrates, yielding wurtzite crystal structure with hexagonal cross section.
The nanowires had core diameters of $\sim140$~nm and were grown to lengths of $\sim10~\mu$m. 
Following semiconductor growth, the substrate stage was cooled to roughly $–30^\circ$C using liquid nitrogen. Al shell was grown \textit{in situ} while rotating the substrate relative to the metal source.
This procedure resulted in a fully surrounding epitaxial Al shell with a clean, oxide-free interface to the InAs core.
The shell thickness was $\sim7$~nm for the nanowire used in device~1 and $\sim24$~nm for device~2.

\textit{Sample preparation:} For device fabrication, individual hybrid nanowires were transferred from the growth substrate onto a degenerately n-doped Si substrate capped with a 200~nm thermal oxide using a home-built manipulator station with a tungsten needle. 
Etch windows, contacts, and gates were defined using standard electron beam lithography (Elionix 7000, 100 kV).  
To improve etch quality, a thin layer of adhesion promoter (Allresist, AR~300-80 new) was applied, followed by a bilayer of EL6 copolymer resist.
The Al shell was selectively removed by immersing the substrate in MF-321 photoresist developer for 75~s at room temperature. 
To contact the Al shell, a single layer of A6 PMMA resist was used for device 1, and an A4/A6 PMMA stack for device 2. 
Prior to metallization (AJA International Inc., Orion), native Al oxides were removed using an Ar-ion milling step (RF ion source, 25~W, 18~mTorr, 8~min for device 1 and 9 min for device 2), followed by deposition of Ti/Al (5/180~nm for device~1 and 5/210~nm for device~2) ohmic contacts.
Normal Ti/Au (5/160 nm for device 1 and 5/180 nm for device 2) ohmic contacts to InAs core were deposited after a gentle Ar-ion milling (RF ion source, 15~W, 18~mTorr, 6.5~min for both devices).
Device~1 additionally featured a Ti/Au (5/150~nm) top gate, patterned using a single layer of A6 PMMA and isolated from the nanowire by an $\sim$8~nm atomic layer deposited (Veeco, Savannah S100) HfO$_x$ dielectric.

%%%%%%%%%%%%%%%%%%%%%% FIG. 5 %%%%%%%%%%%%%%%%%%%%%%%%
\begin{figure}[t!]
\includegraphics[width= \linewidth]{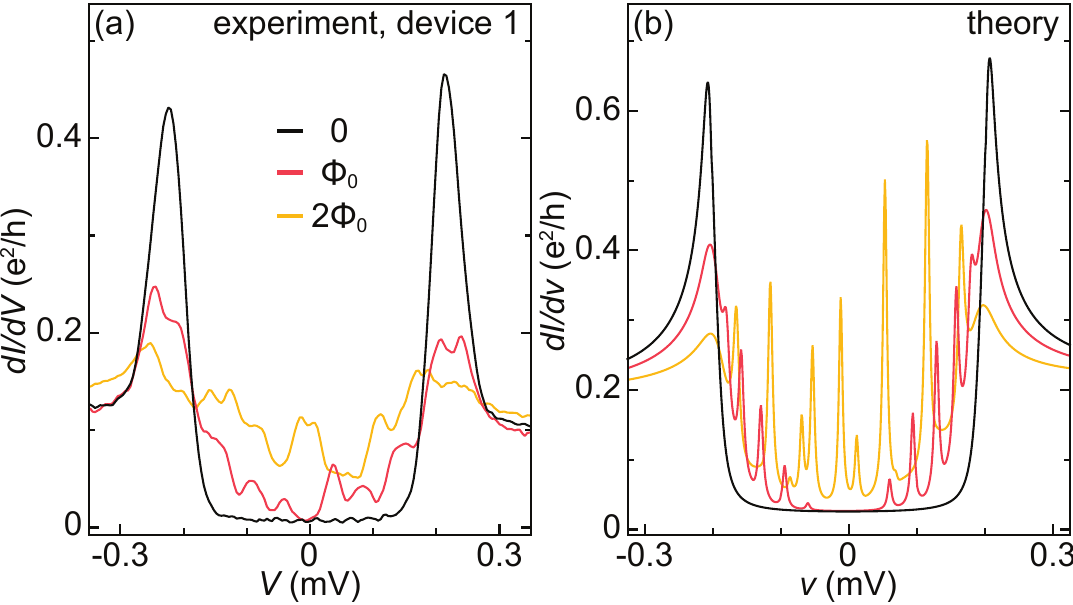}
\caption{\label{emfig:1}
Line cuts of (a) measured differential conductance, $dI/dV$, for device 1 and (b) modeled differential conductance, $dI/dv$, taken from data shown in the main-text Figs.~\ref{fig:2}(a) and~\ref{fig:3}(d) at flux, $\Phi = $ 0, $\Phi_0$, and $2\Phi_0$. 
}
\end{figure}
%%%%%%%%%%%%%%%%%%%%%%%%%%%%%%%%%%%%%%%%%%%%%%%%%%%%%%

\textit{Measurements:} Transport measurements were performed in a dilution refrigerator (Bluefors, XLD-400) with a base temperature of 20~mK, equipped with a three-axis (1, 1, 6)~T vector magnet.
Standard low-frequency lock-in (Stanford Research, SR830) techniques at 127~Hz for device 1 and 131 Hz for device 2 were used to measure differential conductance in a two-terminal geometry.
All dc lines used for measurement and gating were equipped with RF and RC filters (QDevil, QFilter), resulting in a line resistance $R_\mathrm{line} = 6.7$~k$\Omega$.
An ac excitation of $V_\mathrm{ac} = 5~\mu$V was applied to perform voltage-bias spectroscopy in the tunneling regime, where the barrier resistance dominates the total resistance.
The third derivative of the current with respect to voltage, $d^3I/dV^3$, was calculated numerically from the measured differential conductance after applying digital smoothing to suppress numerical noise.

\textit{Line cuts:} Line cuts of measured $dI/dV$ for device 1 [from Figs.~\ref{fig:2}(a)] and modeled $dI/dv$ [from Figs.~\ref{fig:3}(d)] are shown in Fig.~\ref{emfig:1}.
These traces offer a complementary perspective to the colormaps, allowing for a more direct comparison of different features.
We note that the theoretical $dI/dv$ was modeled at $T=0$, whereas the effective electron temperature in the experiment is roughly 40~mK, resulting in broadened spectral features.

\onecolumngrid
\clearpage
\end{document}